\documentclass[pra,twocolumn,showpacs,preprintnumbers,amsmath,amssymb]{revtex4}
\usepackage{epsfig}
\usepackage{color}
\usepackage{graphicx}
\usepackage{dcolumn}
\usepackage{bm}
\newcommand{\bra}[1]{\langle#1|}
\newcommand{\ket}[1]{|#1\rangle}

\providecommand{\openone}{\leavevmode\hbox{\small1\kern-3.8pt\normalsize1}}

\begin{document}

\title{Dynamics of geometric and entropic quantifiers of correlations in open quantum systems}

\author{B. Bellomo,$^1$ R. Lo Franco,$^{1,2}$ and G. Compagno$^1$}
\affiliation{$^1$CNISM and Dipartimento di Fisica, Universit\`a di Palermo - via Archirafi 36, 90123 Palermo, Italy\\
$^2$Centro Siciliano di Fisica Nucleare e di Struttura della Materia (CSFNSM), Viale S. Sofia 64, 95123 Catania, Italy}

\date{\today}

\begin{abstract}
We extend the Hilbert-Schmidt (square norm) distance, previously used to define the geometric quantum discord, to define also geometric quantifiers of total and classical correlations. We then compare the dynamics of geometric and entropic quantifiers of the different kinds of correlations in a non-Markovian open two-qubit system under local dephasing. We find that qualitative differences occur only for quantum discords. This is taken to imply that geometric and entropic discords are not, in general, equivalent in describing the dynamics of quantum correlations. We then show that also geometric and entropic quantifiers of total correlations present qualitative disagreements in the state space. This aspect indicates that the differences found for quantum discord are not attributable to a different separation, introduced by each measure, between the quantum and classical parts of correlations. Finally, we find that the Hilbert-Schmidt distance formally coincides with a symmetrized form of linear relative entropy.
\end{abstract}


\maketitle

\section{Introduction}
To characterize the nature of correlations in a quantum system, and in particular quantifying their quantum and classical parts, is of both fundamental and practical interest. In fact, in certain computational tasks with mixed quantum states quantum speed-up can be achieved using separable (unentangled) states, like in the so-called deterministic quantum computation with one qubit (DQC1) protocol \cite{Knill1998PRL,Lanyon2008PRL}. This speed-up has been linked \cite{Datta2008PRL} to the presence of quantum discord \cite{Zurek2001PRL,Henderson2001JPA}, that quantifies quantum correlations in a bipartite system and was originally defined as the difference between two alternative quantum versions of two classically equivalent expressions of mutual information. Quantum correlations have otherwise been quantified exploiting the idea that the distance between a given state of the system and its closest state without the desired property (e.g., quantum correlations) quantifies that property. The distance is defined by the relative entropy between states and the discord is thus defined as the distance between the system state $\rho$ and its closest classical state $\chi_\rho$ \cite{Modi2010PRL}. The relative-entropy-based (REB) definition of quantum discord does not in general coincide with the original definition of quantum discord but it is connected to the latter by a quantity measuring the distance between two states linked to the state $\rho$ \cite{Modi2010PRL}. Within this approach classical correlations have been then characterized by the distance between the classical state $\chi_\rho$ and its closest product state $\pi_{\chi_\rho}$ while total correlation by the distance between the system state $\rho$ and its closest product state $\pi_\rho$.

The REB quantifiers of correlations (quantum or classical) present the advantage of not being limited, as was the case for the original definition of quantum discord, to bipartite quantum systems but have the drawback that their analytical expressions are known only for certain classes of states \cite{Luo2008PRA,Ali2010PRA} and  require in general considerable numerical minimizations. Thus, a more manageable quantifier of quantum correlations, named geometric quantum discord, has been defined as the Hilbert-Schmidt (square norm) distance between the system state and its closest classical state \cite{dakic2010PRL}. When one studies evolutions of quantum correlations in a two-qubit system locally affected by non-dissipative channels, REB quantum discord may exhibit plateaux during its evolution \cite{mazzola2010PRL,mazzola2010frozen}; differently, in correspondence of these plateaux, geometric quantum discord does not remain constant \cite{lu2010QIC,Xu2010}. These qualitative differences emerge also when the closest classical states entering the two distance measures are the same \cite{Xu2010,bellomo2012PRA}. These findings may be related to the known result in the theory of entanglement that different entanglement measures induce different orderings in the state space \cite{Eisert1999JMO,Virmani2000PLA,Miranowicz2004PRA}. REB and geometric discord have been compared in the state space for two-qubit states and it has been shown that inequivalent ordering also occurs \cite{Adesso2011}. Moreover, an analysis in the state space has recently shown that geometric quantifiers of total, classical and quantum correlations do not satisfy in general closed additivity relations, that instead hold for the entropic quantifiers \cite{bellomo2012PRA}.

The main aim of this paper is to study the dynamics of geometric and entropic quantifiers of the different kinds of correlations (total, classical and quantum) in non-Markovian open quantum systems. To this aim, we consider a specific non-Markovian open two-qubit system under local phase-flip channels. We also compare geometric and entropic quantifiers of the different kinds of correlations in the state space moving inside the class of Bell-diagonal states. Finally, we observe that the Hilbert-Schmidt distance formally coincides with a symmetrized form of linear relative entropy.

\section{Geometric quantifiers of total and classical correlations}
Before introducing the geometric quantifiers of total and classical correlations, we briefly review the correlation quantifiers based on relative entropy as measure of distance between states. Relative entropy quantifies the distinguishability between two states $\rho$ and $\sigma$ and is defined as
\begin{equation}\label{relativeentropy}
S(\rho\|\sigma)=-\mathrm{Tr}(\rho\log_2\sigma)-S(\rho),
\end{equation}
where $S(\rho)=-\mathrm{Tr}(\rho\log_2\rho)$ is the von Neumann entropy \cite{vedral2002RevModPhys}. $S(\rho\|\sigma)$ is adopted as a measure of the distance between the states $\rho$ and $\sigma$ even if it is asymmetric with respect to the exchange $\rho\leftrightarrow\sigma$ and is thus a pseudo-distance.
In this view, total correlations (quantum mutual information) $T$, discord $D$ and classical correlations $C$ have been defined as \cite{Modi2010PRL}
\begin{equation}\label{quantifiers}
T(\rho)\equiv S(\rho\|\pi_\rho),\
D(\rho)\equiv S(\rho\|\chi_\rho),\
C(\rho)\equiv S(\chi_\rho\|\pi_{\chi_\rho}),
\end{equation}
where $\pi_\rho$ and $\chi_\rho$ are respectively the product state and the classical state closest to $\rho$, while $\pi_{\chi_\rho}$ is the product state closest to $\chi_\rho$. We refer to $T$, $D$ and $C$ as REB quantifiers. As already said, $D(\rho)$ does not coincide, in general, with the original definition of quantum discord $\delta(\rho)\equiv\mathrm{min}_{\Pi^A}\{I(\rho)-I[\Pi^A(\rho)]\}$ \cite{Luo2010PRA}, where $I(\rho)$ is the quantum mutual information, the minimum is over von Neumann measurements $\Pi^A$ and $\Pi^A(\rho)$ is the classical state resulting after projective measurement on $A$. However, $D(\rho)$ and $\delta(\rho)$, for a two-qubit system, result equal for Bell-diagonal states \cite{Modi2010PRL}. For REB distance measure the closest product state $\pi_\rho$ is the product of the marginals of $\rho$ ($\pi_\rho=\rho_A\otimes\rho_B$), while $\chi_\rho$ (and therefore also $D$ and $C$) is analytically obtainable only within certain classes of states, even for a two-qubit system \cite{Modi2010PRL}.

A more manageable distance measure between states is the Hilbert-Schmidt (square-norm) distance, by which the geometric quantum discord has been defined as \cite{dakic2010PRL}
\begin{equation}\label{geometricdiscord}
   D_\mathrm{g}(\rho)=\mathrm{Tr}(\rho-\chi_\rho)^2=\|\rho-\chi_\rho\|^2,
\end{equation}
where $\chi_\rho$ is the classical state closest to $\rho$. This geometric discord, differently from the entropic discord, can be analytically evaluated for an arbitrary two-qubit state finding the explicit expression of its closest classical state \cite{dakic2010PRL}. Notice that the geometric quantum discord $D_\mathrm{g}(\rho)$ for a bipartite system is equivalent to the one obtained as the minimum distance, measured by the square norm, between $\rho$ and a classical state $\Pi^A(\rho)$ resulting from $\rho$ after projective measurement on $A$, that is $D_\mathrm{g}(\rho)=\mathrm{min}_{\Pi^A}\|\rho-\Pi^A(\rho)\|^2$ \cite{Luo2010PRA}.
Quantifiers of total and classical correlations, analogously to the geometric quantum discord, can be now introduced using the Hilbert-Schmidt distance as a measure of the distance between states, provided that one finds the product states $\pi_\rho$ and $\pi_{\chi_\rho}$ closest to $\rho$ and $\chi_\rho$, respectively. We define geometric quantifiers of total and classical correlations as
\begin{equation}\label{geometric total and classical}
    T_\mathrm{g}(\rho)\equiv \|\rho-\pi_\rho\|^2,\quad C_\mathrm{g}(\rho)\equiv \|\chi_\rho-\pi_{\chi_\rho}\|^2.
\end{equation}
These quantifiers do not satisfy in general an additivity relation, that is instead satisfied in the case of Bell-diagonal states for which $D_\mathrm{g}(\rho)=T_\mathrm{g}(\rho)-C_\mathrm{g}(\rho)$ \cite{bellomo2012PRA}. It is also shown in appendix A that, for this class of states, the closest product state is given by the product of its marginals, $\pi_\rho=\rho_A\otimes\rho_B$, exactly as happens for the REB distance measure.

\subsection{Expressions of the correlation quantifiers for Bell-diagonal states}
We now give the explicit expressions of REB and geometric quantifiers discussed above for the class of Bell-diagonal states. In the Bell-state basis and in the Bloch state representation, a Bell-diagonal state is written, respectively, as
\begin{equation}\label{initialBelldiagonalstate}
\rho^\mathrm{B}=\sum_{i,r}\lambda_i^r\ket{i_r}\bra{i_r},\quad
\rho^\mathrm{B}=[\mathbb{I}^A\otimes\mathbb{I}^B+\sum_{j=1}^3c_{j}\sigma_j\otimes\sigma_j]/4,
\end{equation}
where $i=1,2$, $r=\pm$, the coefficients $c_j$'s and $\lambda_i^r$ are real and where we have indicated with $\ket{1_\pm}\equiv(\ket{01}\pm\ket{10})/\sqrt{2}$ the one-excitation Bell states and with $\ket{2_\pm}\equiv(\ket{00}\pm\ket{11})/\sqrt{2}$ the two-excitation Bell states. The states $\rho^\mathrm{B}$ include the well-known Werner states \cite{bellomo2008PRA,bellomo2009ASL} and are entangled if the largest $\lambda_i^r>1/2$ \cite{Modi2010PRL}. The product state closest to $\rho^\mathrm{B}$ is equal to the normalized $4\times4$ identity, $\pi_{\rho^\mathrm{B}}=\mathbb{I}/4=(\mathbb{I}^A/2)\otimes(\mathbb{I}^B/2)$ for both distance measures (relative entropy and square norm, see Appendix A). The analytic expression of the closest classical state when the distance is measured by relative entropy is known \cite{Modi2010PRL}. For Bell-diagonal states, the REB quantifiers of Eq.~(\ref{quantifiers}) are 
\begin{eqnarray}\label{REBquantifiers}
D(\rho)&=&T(\rho)-C(\rho),\nonumber\\ 
T(\rho)&=&2+\sum_{i,r}\lambda_i^r\log_2\lambda_i^r, \quad (i=1,2; r=\pm)\nonumber\\
C(\rho)&=&\sum_{i=1}^2 \frac{1+(-1)^ic}{2}\mathrm{log}[1+(-1)^ic],
\end{eqnarray}
where $c\equiv\mathrm{max}\{|c_1|,|c_2|,|c_3|\}$ \cite{Luo2008PRA}. 

On the other hand, the classical state closest to $\rho^\mathrm{B}$ according to the Hilbert-Schmidt distance is still a Bell-diagonal state and results to be
\begin{equation}\label{classicalstate}
\chi_{\rho^\mathrm{B}}=\left[\mathbb{I}^A\otimes\mathbb{I}^B+c_k\sigma_k\otimes\sigma_k\right]/4,
\end{equation}
where $c_k$ is the one among the coefficients $c_1,c_2,c_3$ such that $|c_k|=c$.  $\chi_{\rho^\mathrm{B}}$ coincides with that obtained using the REB distance measure. The expressions of the geometric discord of Eq.~(\ref{geometricdiscord}) and of the geometric quantifiers of total and classical correlations of Eq.~(\ref{geometric total and classical}) are then given by 
\begin{eqnarray}\label{Bellgeometricquantifiers}
D_\mathrm{g}(\rho)&=&[c_1^2+c_2^2+c_3^2-c^2]/4,\nonumber\\
T_\mathrm{g}(\rho)&=&[c_1^2+c_2^2+c_3^2]/4,\nonumber\\
C_\mathrm{g}(\rho)&=&c^2/4. 
\end{eqnarray}
Notice that the closest classical state $\chi_{\rho^\mathrm{B}}$ above is symmetric under exchange of subsystems $A$, $B$, thus it has the same value of left and right discord \cite{dakic2010PRL}. Because $D_\mathrm{g}$ indicates the left discord and $D_\mathrm{g}(\chi_{\rho^\mathrm{B}})=0$, the state $\chi_{\rho^\mathrm{B}}$ of Eq.~(\ref{classicalstate}) is a classical-classical state \cite{dakic2010PRL}. Notice that for Bell-diagonal states one gets $\mathrm{Tr}(\rho^\mathrm{B}\chi_{\rho^\mathrm{B}})=\mathrm{Tr}(\chi_{\rho^\mathrm{B}}^2)$, so that $D_\mathrm{g}\equiv\mathrm{Tr}(\rho^\mathrm{B}-\chi_{\rho^\mathrm{B}})^2=\mathrm{Tr}[(\rho^\mathrm{B})^2-\chi_{\rho^\mathrm{B}}^2]$. We also point out that expressions of the geometric correlation quantifiers for the more general class of X two-qubit states have been recently reported in the case the closest classical states involved have a classical-quantum form \cite{bellomo2012PRA}. 

In the following, we shall compare these geometric quantifiers with the corresponding REB correlation quantifiers both in a dynamical contexts and in the state space.

\section{Comparisons between correlation quantifiers}
In this section, we compare quantum discord and total correlations quantified according to the two distance measures with the aim to make evident possible different qualitative behaviors in the correlation quantifiers. Due to the fact that the square norm distance is not normalized, $D_\mathrm{g}$ is not normalized to one (its maximum value is 1/2 for two-qubit states) \cite{Adesso2011,DakicZeilingerl2012arXiv}. We therefore shall consider the normalized forms $2D_\mathrm{g}$ and $2T_\mathrm{g}$ as proper quantifiers for a comparison with REB quantum discord $D$ and REB total correlations $T$.

\subsection{Comparison between REB discord and geometric discord in a dynamical example}
\begin{figure}
\begin{center}
{\includegraphics[width=0.4\textwidth]{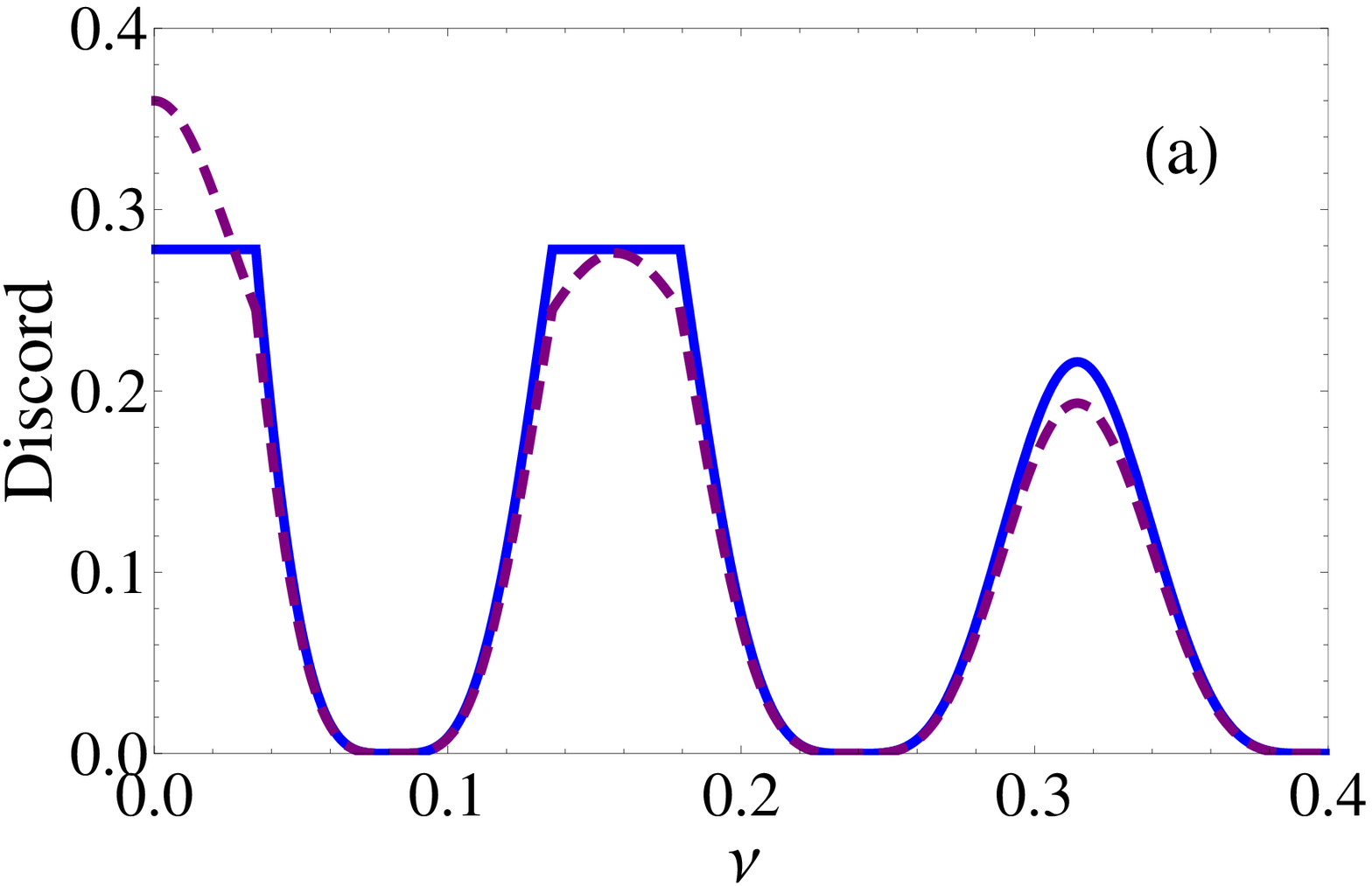}\vspace{0.5 cm}
\includegraphics[width=0.4\textwidth]{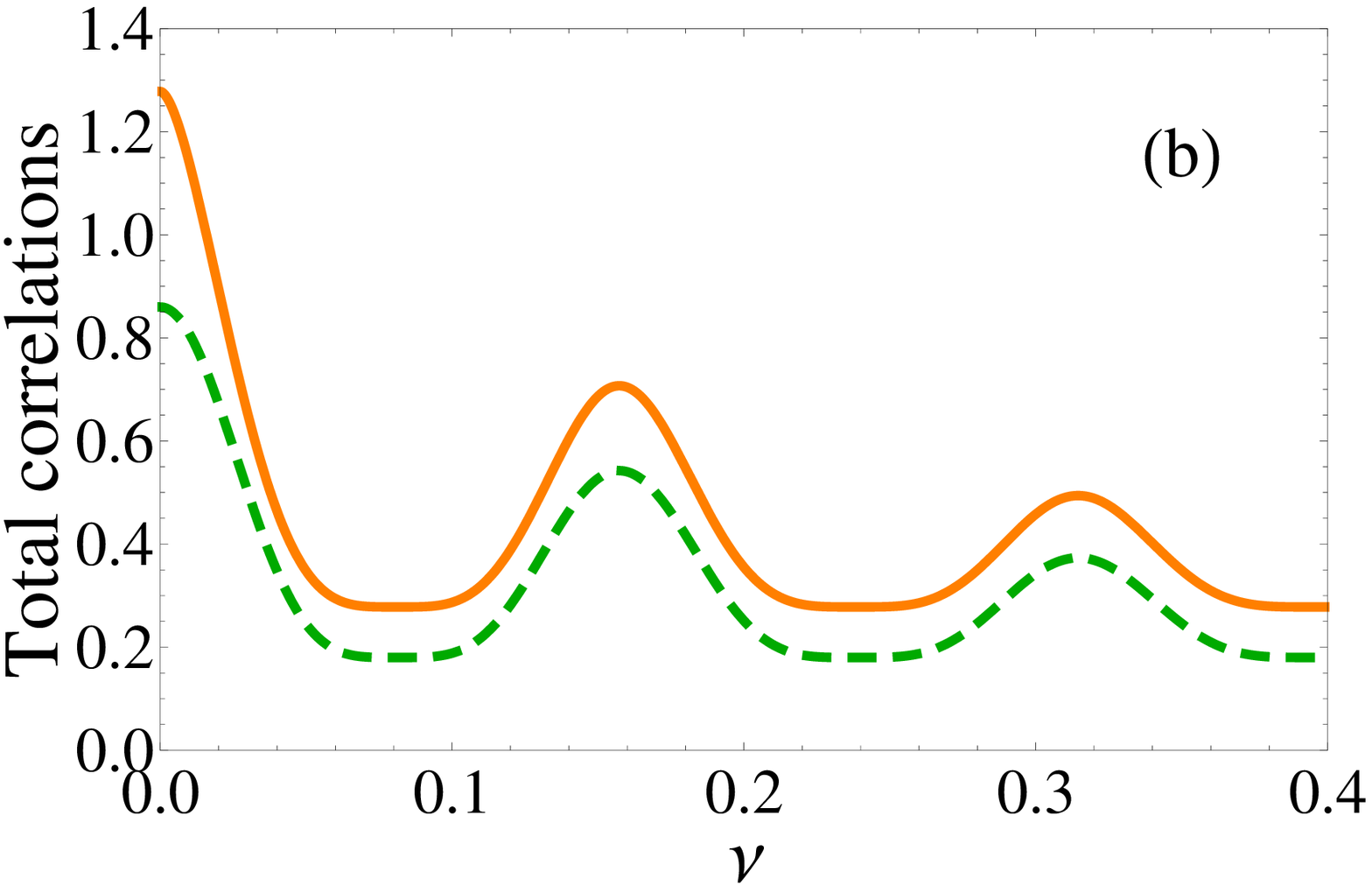}}
\end{center}
\caption{\label{fig1:REBconstant}\footnotesize (Color online) Panel (a): REB quantum discord $D$ (blue solid line) and normalized geometric quantum discord $2D_\mathrm{g}$ (purple dashed line) versus $\nu=t/2\tau$, with $\tau=5$s and $|\alpha|=1$s$^{-1}$, for an initial Bell-diagonal state. Panel (b): REB total correlations $T$ (orange solid line) and normalized geometric total correlations $2T_\mathrm{g}$ (green dashed line) versus $\nu$. The values of initial coefficients are: $c_1(0)=1$, $c_2(0)=-0.6 $ and $c_3(0)=0.6$ ($\lambda_1^+(0)=0.2$, $\lambda_1^-(0)=\lambda_2^-(0)=0$, $\lambda_2^+(0)=0.8$).}
\end{figure}

We firstly take two noninteracting qubits under local identical phase flip channels \cite{mazzola2010frozen}. Each qubit is subject to a time-dependent phenomenological Hamiltonian $H(t)=\hbar\Gamma(t)\sigma_z$, where $\sigma_z$ is a Pauli operator and $\Gamma(t)=\alpha n(t)$ where $\alpha$ is a coin-flip random variable taking the values $\pm |\alpha|$ while $n(t)$ is a random variable having a Poisson distribution with mean value equal to the dimensionless time $\nu=t/2\tau$. This system is characterized by a non-Markovian dynamics of the two-qubit state that, if initially in Bell-diagonal form, remains of this kind during the dynamics and correspondingly the three coefficients $c_j(t)$ of Eq.(\ref{initialBelldiagonalstate}) evolve as
\begin{equation}\label{timedependentc}
  c_{j'}(t)=c_{j'}(0){\Lambda(\nu)}^2, \quad c_3(t)=c_3(0),
\end{equation}
where $j'=1,2$ and $\Lambda(\nu)=\mathrm{e}^{-\nu}[\cos(\mu \nu)+\sin(\mu \nu)/\mu]$ with $\mu=\sqrt{(4\alpha\tau)^2-1}$. Using Eq.~(\ref{timedependentc}), all the quantifiers of Eqs.~(\ref{REBquantifiers}) and (\ref{Bellgeometricquantifiers}) can be analytically computed and all the relevant closest states for both distance measures result the same at any time $t$. In particular, $\pi_{\rho(t)}$ and $\pi_{\chi_\rho(t)}$ remain equal to $\mathbb{I}/4$ while $\chi_{\rho^\mathrm{B}(t)}=\left[\mathbb{I}\otimes\mathbb{I}+c_k(t)\sigma_k\otimes\sigma_k\right]/4$. We also notice that the closest classical state $\chi_{\rho^\mathrm{B}(t)}$ of Eq.~(\ref{classicalstate}) is frozen during the time intervals when $|c_3(t)|>|c_1(t)|,|c_2(t)|$, being $c_k(t)=c_3(t)=c_3(0)$. In Fig.~\ref{fig1:REBconstant}(a) REB discord $D$ and normalized geometric discord $2D_\mathrm{g}$ are plotted as a function of the dimensionless time $\nu$ for a given initial Bell-diagonal state.
This plot shows that there are time regions when $D$ is constant while $D_\mathrm{g}$ decreases or increases. The reason of the different behavior of $D$ and $D_\mathrm{g}$ is connected to the property that the closest classical state varies with time when $|c_1(t)|>|c_2(t)|,|c_3(t)|$. Therefore, its distance from the system state $\rho(t)$ is in general expected to explicitly depend on time. This happens for geometric discord $D_\mathrm{g}$ while, for REB discord $D$, this time-dependence disappears as a consequence of the additivity of logarithm. Notice that an analogous time behavior of REB quantum discord is found when the two qubits are locally subject to random external fields \cite{lofranco2012PRA}.

\begin{figure}
\begin{center}
{\includegraphics[width=0.4\textwidth]{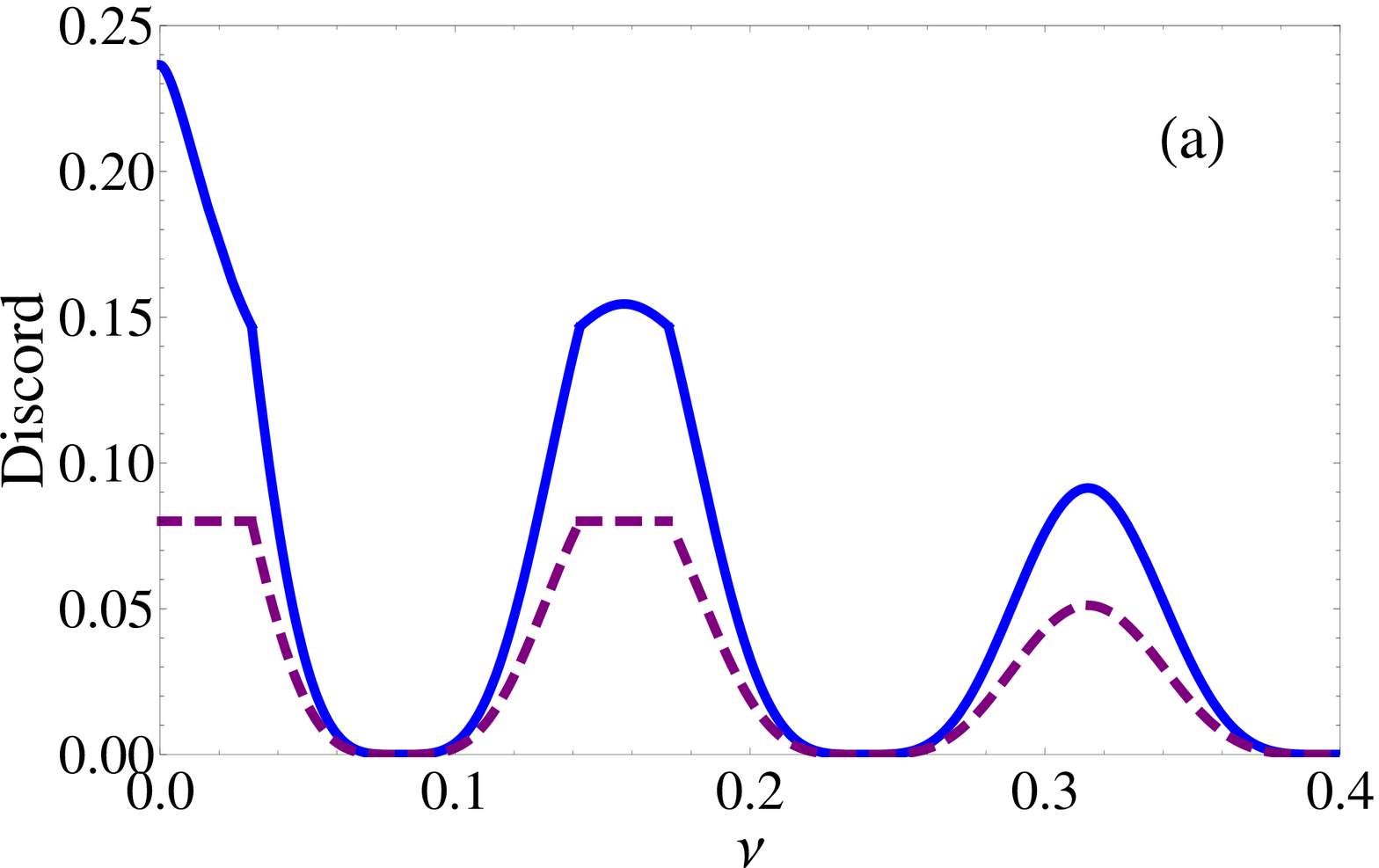}\vspace{0.5 cm}
\includegraphics[width=0.4\textwidth]{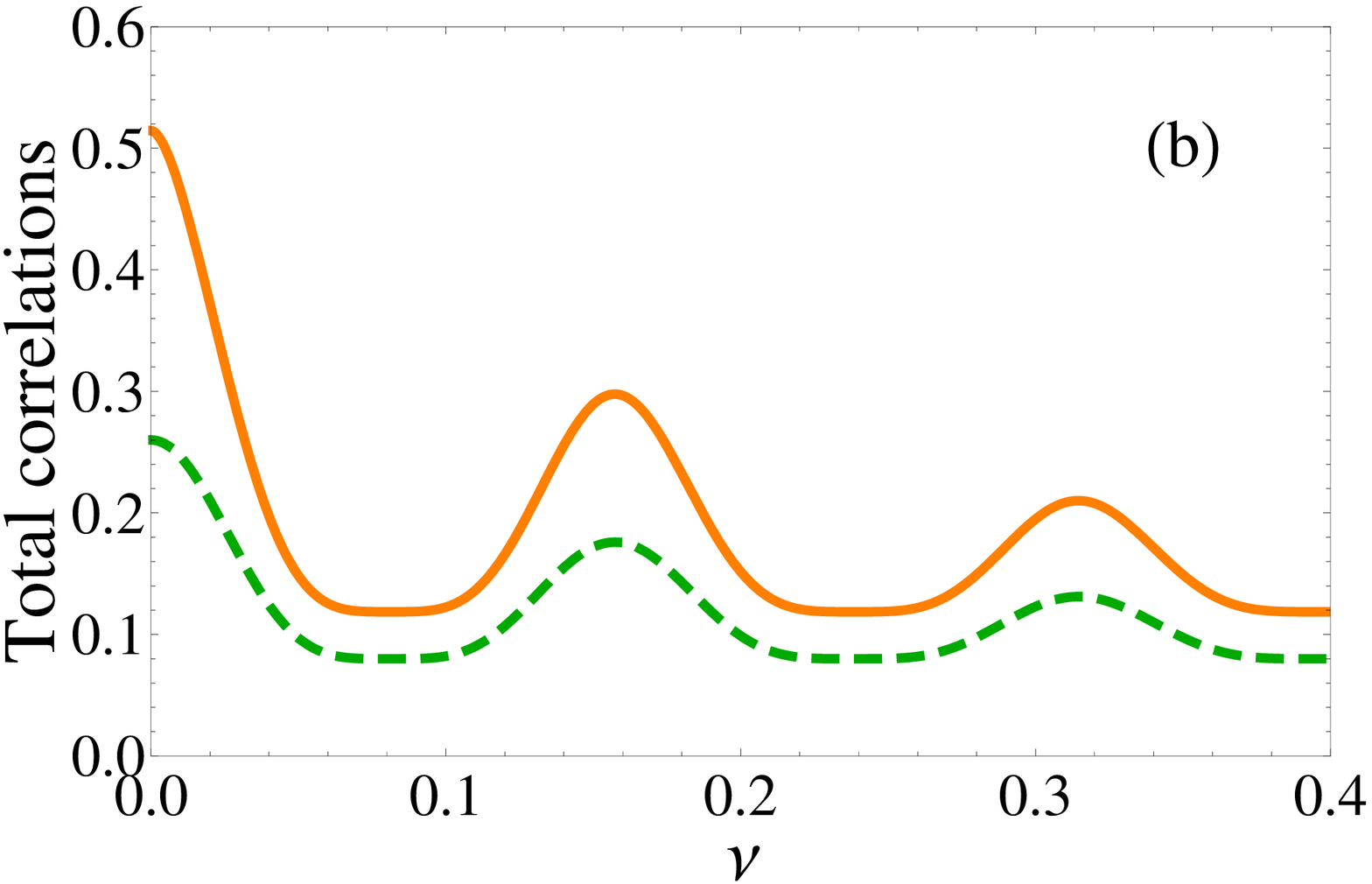}}
\end{center}
\caption{\label{fig2:geometricconstant}\footnotesize (Color online) Panel (a): REB quantum discord $D$ (blue solid line) and normalized geometric quantum discord $2D_\mathrm{g}$ (purple dashed line) versus $\nu$, for an initial Bell-diagonal state with $c_1(0)=0.6$, $c_2(0)=0 $ and $c_3(0)=0.4$ ($\lambda_1^+(0)=0.3$, $\lambda_1^-(0)=0$, $\lambda_2^+(0)=0.5$, $\lambda_2^-(0)=0.2$). Panel (b): REB total correlations $T$ (orange solid line) and normalized geometric total correlations $2T_\mathrm{g}$ (green dashed line) versus $\nu$. Values of other parameters for both panels: $\tau=5$s, $|\alpha|=1$s$^{-1}$.}
\end{figure}

The plot of Fig.~\ref{fig2:geometricconstant}(a) then shows that, for a different initial Bell-diagonal state, a behavior opposite to the previous one may occur: there are time regions when $D_\mathrm{g}$ is constant while $D$ decreases or increases. The behavior of $D$ and $D_\mathrm{g}$ in panels (a) occurs for initial states with $c_{2(1)}(0)=0$ and $|c_{1(2)}(0)|>|c_3(0)|$, so that one has $D_\mathrm{g}=\frac{1}{4}[c_3(0)^2]$. 
At this point, the different qualitative time behaviors between $D$ and $D_\mathrm{g}$ can be considered as due to the different choice of distance measure.

In Figures~\ref{fig1:REBconstant}(b) and \ref{fig2:geometricconstant}(b) total correlation quantifiers are plotted. It is displayed that, in spite of the differences between $D$ and $D_\mathrm{g}$, the dynamics of total correlations qualitatively behave in a similar way, that is they are increasing or decreasing in the same time regions, for both distance measures. The quantifiers of classical correlations $C$ and $C_\mathrm{g}$ also exhibit a dynamical behavior in agreement to each other. This fact could induce to reckon that different qualitative behaviors, found for quantum discords, emerge as a consequence of trying to distinguish the quantum and classical part of the total correlations in a composite system. In the following we shall investigate this aspect by considering particular physical states in the state space.

\subsection{Comparison between REB and geometric total correlations in the space state}
We compare total correlations based on the two distance measures for a particular subset of Bell-diagonal states in the space state. Our aim is just to give an example of different physical states for which the amounts of REB and geometric total correlations give disagreeing results.

\begin{figure}
\begin{center}
{\includegraphics[width=0.42\textwidth]{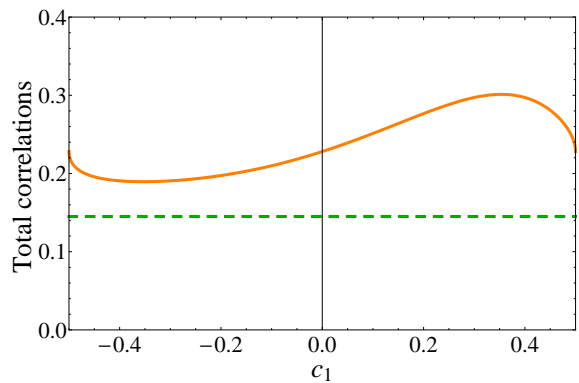}}
\end{center}
\caption{\label{fig3:totalcorrelations}\footnotesize (Color online) REB total correlations $T$ (orange solid line) and geometric total correlations $T_\mathrm{g}$ (green dashed line), in the space state, as functions of $c_1$, with $c_2=\sqrt{0.25-c_1^2}$ and $c_3=0.2$.}
\end{figure}

We consider Bell-diagonal states, as defined in Eq.~(\ref{initialBelldiagonalstate}), such that geometric total correlations $T_\mathrm{g}$ of Eq.~(\ref{Bellgeometricquantifiers}) have the same value: this means that $\sum_{i=1}^3c_i^2=\mathrm{constant}$ for all of them. The coefficients $c_i$ must be also such as to satisfy the conditions of physical state for the elements of the density matrix $\rho^\mathrm{B}$ written in the computational basis $\mathcal{B}=\{\ket{1}\equiv\ket{11},
\ket{2}\equiv\ket{10}, \ket{3}\equiv\ket{01}, \ket{4}\equiv\ket{00} \}$, that is \cite{nielsenchuang} $|\rho_{14}|^2\leq\rho_{11}\rho_{44}$ and $|\rho_{23}|^2\leq\rho_{22}\rho_{33}$. The relations among coefficients $c_i$ and density matrix elements are
\begin{eqnarray}\label{cdensitymatrix}
\rho_{11}&=&\rho_{44}=(1+c_3)/4,\quad \rho_{22}=\rho_{33}=(1-c_3)/4,\nonumber\\ 
\rho_{14}&=&(c_1-c_2)/4,\quad \rho_{23}=(c_1+c_2)/4.
\end{eqnarray}
We then choose $c_3=0.2$, $c_2=\sqrt{0.25-c_1^2}$ and let $c_1$ vary from $-0.5$ to $0.5$. This choice satisfies the constraint of physical state. The plot of REB quantifier $T$ and normalized geometric quantifier $2T_\mathrm{g}$ of total correlations is given in Fig.~\ref{fig3:totalcorrelations}. It is seen that states with the same total correlations, as measured by Hilbert-Schmidt distance, exhibit different values of total correlations if measured by REB distance. In this case the differences observed seems to be more striking because they involve the totality of correlations present in the system state and not only a part of them, as happened in the dynamical case treated above.

\begin{figure}
\begin{center}
{\includegraphics[width=0.23\textwidth]{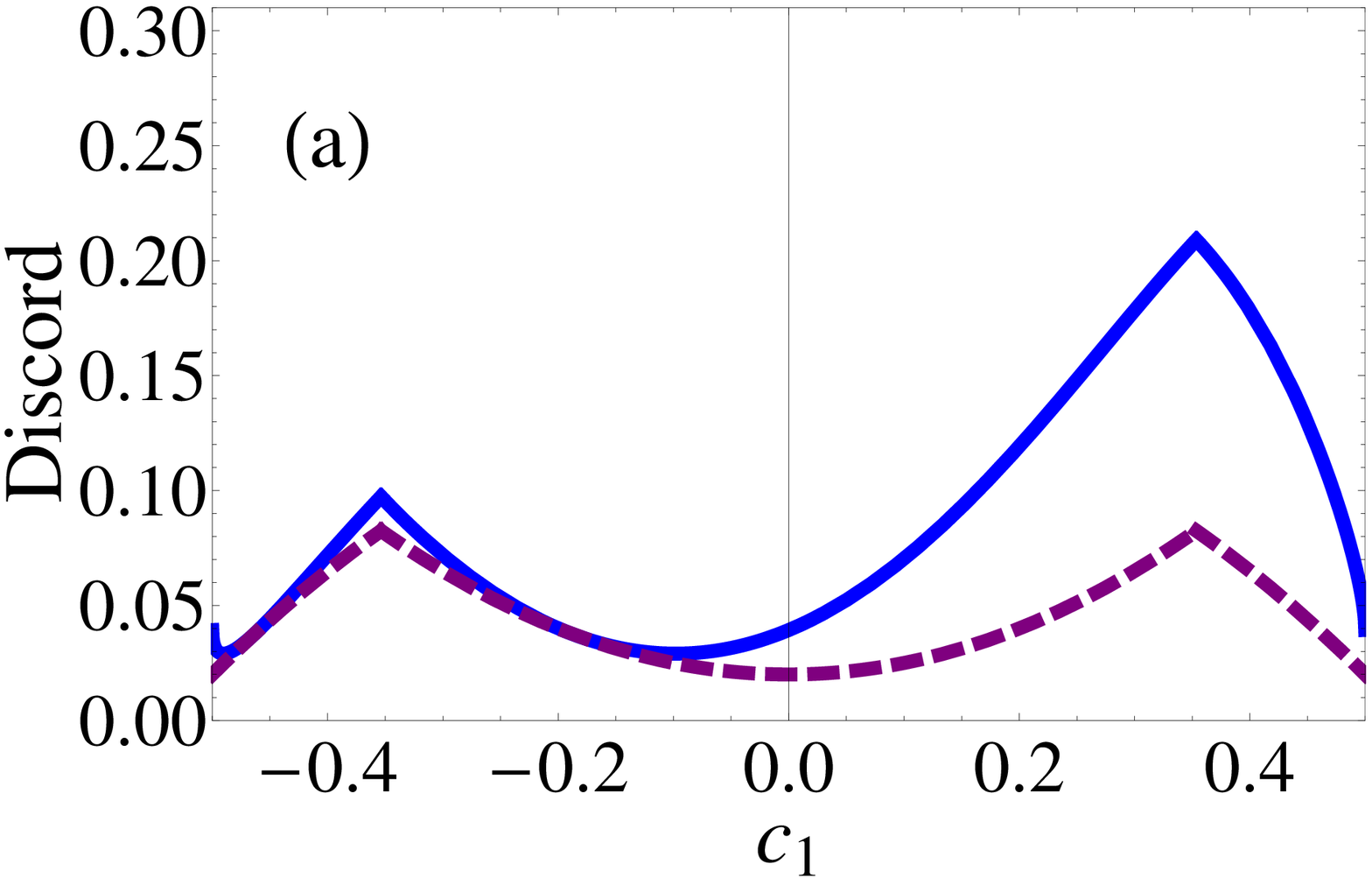}\hspace{0.3 cm}
\includegraphics[width=0.23\textwidth]{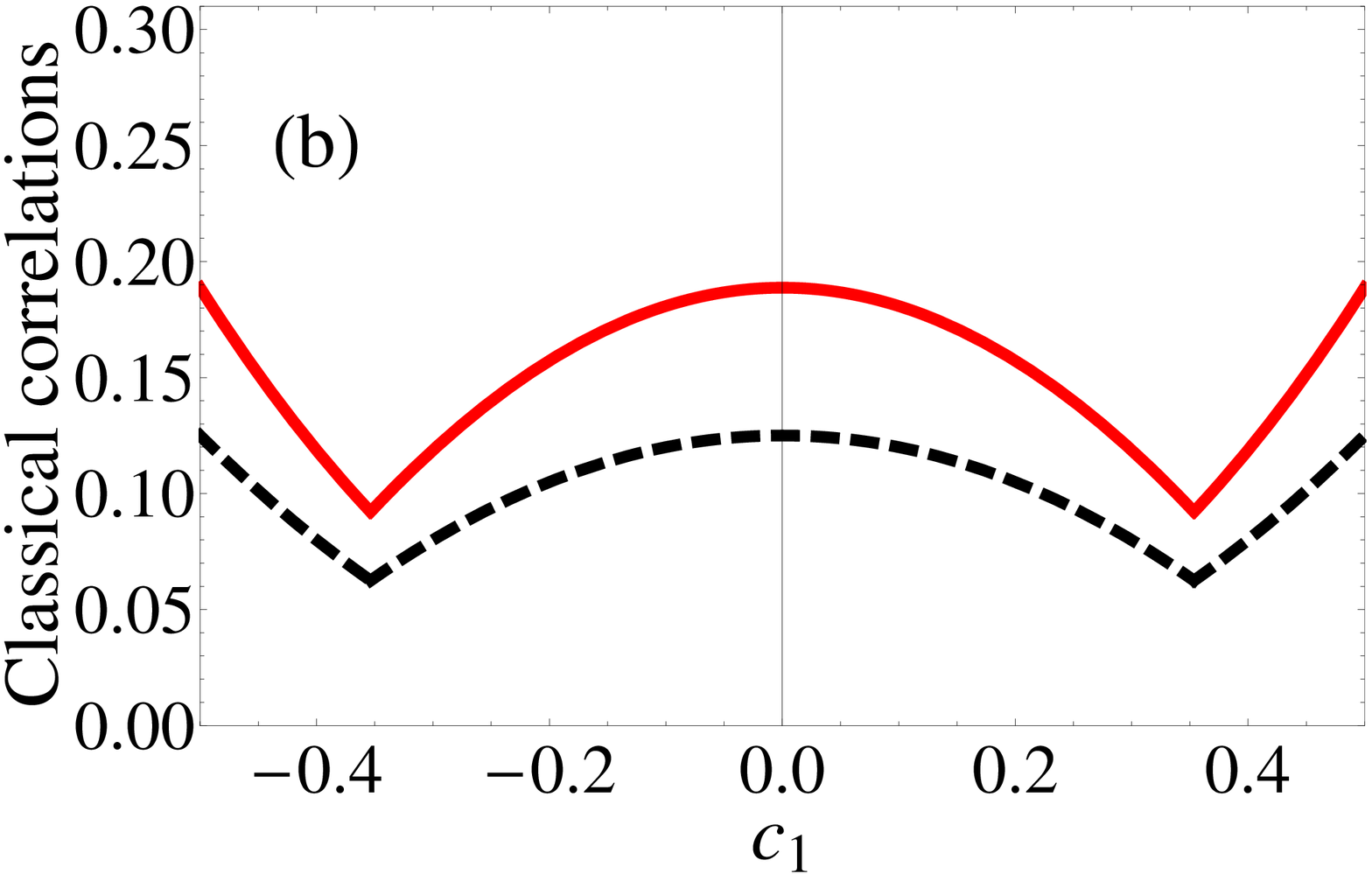}}
\end{center}
\caption{\label{fig4:DCspacestate}\footnotesize (Color online) Panel (a): REB discord $D$ (blue solid line) and normalized geometric discord $2D_\mathrm{g}$ (purple dashed line), in the space state, as functions of $c_1$. Panel (b): REB classical correlations $C$ (red solid line) and normalized geometric classical correlations $2C_\mathrm{g}$ (black dashed line) as functions of $c_1$. The other coefficients are $c_2=\sqrt{0.25-c_1^2}$ and $c_3=0.2$}
\end{figure}

To conclude our analysis we also plot in Fig.~\ref{fig4:DCspacestate} quantum discords $D$, $2D_\mathrm{g}$ and classical correlation quantifiers $C,2C_\mathrm{g}$ for the same Bell-diagonal states above. It is displayed, as known \cite{Adesso2011}, as quantum correlations present inversion of ordering for some states when measured by REB quantum discord or geometric discord (compare, for example, states with $c_1=-0.3$ and $c_1=0.2$ of Fig.~\ref{fig4:DCspacestate}(a)). That is, we have two states $\rho_1$ (corresponding to $c_1=-0.3$), $\rho_2$ (corresponding to $c_2=0.2$) for which $D(\rho_1)<D(\rho_2)$ while $D_\mathrm{g}(\rho_1)>D_\mathrm{g}(\rho_2)$. It is instead seen from Fig.~\ref{fig4:DCspacestate}(b) that there is a qualitative agreement between the two quantifiers of classical correlations, that is if  $C(\rho_1)<C(\rho_2)$ then also $C_\mathrm{g}(\rho_1)<C_\mathrm{g}(\rho_2)$. According to these results and to those found in the previous dynamical example, we have indications that the quantifiers of classical correlations $C,C_\mathrm{g}$ seem to always behave in a qualitative similar way.

\section{Symmetrized linear relative entropy and Hilbert-Schmidt distance}
The mixedness of a state $\rho$ can be quantified, besides by the von Neumann entropy, by the linear entropy $S_L(\rho)=1-\mathrm{Tr}(\rho^2)$ \cite{nielsenchuang}. We point out that $S_L(\rho)$ can be formally obtained from von Neumann entropy $S(\rho)$ by approximating the logarithms with the first terms $(\rho-\mathbb{I})$ in the Mercator series ($\mathbb{I}$ is the identity matrix). One is thus led to apply the same formal approximation directly in the definition of relative entropy $S(\rho\|\sigma)$ of Eq.~(\ref{relativeentropy}). One thus gets the expression of linear relative entropy as
\begin{equation}\label{linear relative entropy}
   S_L(\rho\|\sigma)=\mathrm{Tr}[\rho(\rho-\sigma)].
\end{equation}
$S_L(\rho\|\sigma)$ shares with the relative entropy a lack of symmetry with respect to the exchange $\rho\leftrightarrow\sigma$ and can be negative.
When $\rho$ is the completely mixed state of a $N$-partite system ($\mathbb{I}/N$), it is straightforward to show that the linear relative entropy above is always zero independently of $\sigma$. This property makes the direct use of $S_L(\rho\|\sigma)$ of Eq.~(\ref{linear relative entropy}) unsuitable as a measure of the distinguishability between two arbitrary states. 

Nevertheless, differently from what happens for relative entropy \cite{Eisert2003JPA}, the quantity obtained from $S_L(\rho\|\sigma)$ by exchanging $\rho\leftrightarrow\sigma$ is well defined. It is therefore possible to construct the two quantities $S_L^a(\rho,\sigma)\equiv S_L(\rho\|\sigma)-S_L(\sigma\|\rho)$ and $S_L^s(\rho,\sigma)\equiv S_L(\rho\|\sigma)+S_L(\sigma\|\rho)$ that are, respectively, the anti-symmetrized and symmetrized forms of linear relative entropy. The anti-symmetrized form $S_L^a(\rho,\sigma)$, when $\sigma=\rho_A\otimes\rho_B$, is seen to coincide with the quantum linear mutual information $I_L(\rho)=S_L(\rho_A\otimes\rho_B)-S_L(\rho)$ introduced in Ref.~\cite{Angelo2004PhysA}. We observe that, while for Bell-diagonal states the geometric quantifier of total correlations of Eq.~(\ref{geometric total and classical}) is $T_\mathrm{g}(\rho)=I_L(\rho)$, in general $T_\mathrm{g}(\rho)$ is not equal to $I_L(\rho)$. 

The symmetrized form $S_L^s(\rho,\sigma)$ instead results to be
\begin{equation}\label{SLRE}
    S_L^s(\rho,\sigma)=\mathrm{Tr}(\rho-\sigma)^2=\|\rho-\sigma\|^2,
\end{equation}
that formally coincides with the Hilbert-Schmidt distance. When $\sigma$ is the closest classical state $\chi_\rho$, we have $S_L^s(\rho,\chi_\rho)=D_\mathrm{g}(\rho)$: this way, the entropic and geometric quantum discords can be both related to entropy measures. This result could open the way to further investigations concerning the connection between quantifiers of correlations based on relative entropy and Hilbert-Schmidt distance.

\section{Conclusions}
We have used Hilbert-Schmidt distance to define geometric quantifiers of classical and total correlations. We have then compared, within an open two-qubit system subject to a suitable phase-damping evolution, the dynamics of entropic (REB) and geometric quantifiers of correlations. We have found that, by appropriately changing the initial state, not only geometric discord may vary in correspondence of constant REB discord but also the opposite may occur, that is there are time regions when REB discord may vary in correspondence of constant geometric discord. Quantifiers of total and classical correlations instead behave in a qualitative similar way. We have then found that total correlations, as quantified by REB distance and Hilbert-Schmidt distance, present qualitative differences in the state space for physical states within the class of Bell-diagonal states. For instance, two-qubit states having the same amount of total correlations according to the Hilbert-Schmidt distance exhibit different amounts of total correlations if measured by REB distance. On the other hand quantifiers of classical correlations, defined by the two distance measures here considered, have a qualitatively similar behavior both dynamically and in the state space. We have finally shown that the Hilbert-Schmidt (square norm) distance formally coincides with the symmetrized form of linear relative entropy. 

These results point out that quantifiers of a specific kind of correlations, based on different distance measures, exhibit not only quantitative but also qualitative differences. For example, while a quantifier of a specific kind of correlations, based on a given distance measure, has a dynamical behavior (e.g., constant or decreasing), another quantifier behaves differently (e.g., increasing). These findings seem to have a counterpart in the relativity of entanglement measures as result of physical processes \cite{Miranowicz2004JPA}: in this case however the relativity of measures shows up when one compares two different states as they evolve from different initial conditions. Here, instead, different qualitative behaviors appear in the evolution of quantum correlation quantifiers that are used to represent physical dynamics of the same kind of correlations present during the evolution of a single state. As a further point, the relationship among geometric total, quantum, and classical correlations has been investigated finding that they do not satisfy, in general, a closed additivity relation \cite{bellomo2012PRA}, as happens instead for REB correlation quantifiers \cite{Modi2010PRL}. Entropic and geometric quantum discords could thus individuate themselves genuinely inequivalent characterizations of nonclassical correlations, as also appears to be corroborated by recent analyses \cite{adesso2012arXiv,piani2012arXiv}. The above results indicate that appropriate quantification of the physical dynamics of correlations present in an open quantum state can be considered not yet completely settled. 

A possible way to overcome these issues could be to link correlation quantifiers to operational tasks. In this context, REB quantum discord has been proposed as the resource to enhance computation \cite{Braunstein1999PRL,Datta2005PRA}, but its relation to the computational speed-up is still not completely clear \cite{dakic2010PRL,Ferraro2010PRA}. A relation of REB quantum discord with quantum communication has also been pointed out but only in few particular cases, for example in local broadcasting \cite{Piani2008PRL} and quantum state merging \cite{Cavalcanti2011PRA,Madhok2011PRA}. A presumably significant step forward has been recently done in the operational interpretation of geometric (SLRE-based) quantum discord. In this interpretation, geometric quantum discord results to be the optimal resource for remote quantum state preparation (a variant of quantum teleportation protocol) \cite{DakicZeilingerl2012arXiv,adesso2012arXiv}.

\begin{acknowledgments}
B.B. and R.L.F. would like to thank Dariusz Chru\'{s}ci\'{n}ski and Bruno Leggio for useful discussions.
\end{acknowledgments}

\appendix
\section{}
Here we show that, in the square norm distance, the product state closest to a Bell-diagonal state $\rho^\mathrm{B}$ of Eq.~(\ref{initialBelldiagonalstate}) is given by the product of its marginals. To this purpose, we consider arbitrary states of qubits $A$ and $B$, respectively, 
\begin{equation}
\tilde{\rho}_A=\frac{1}{2}[\mathbb{I}^A+\sum_i a_i\sigma_i], \quad
\tilde{\rho}_B=\frac{1}{2}[\mathbb{I}^B+\sum_i b_i\sigma_i].
\end{equation}
Their product is $\tilde{\rho}_A\otimes \tilde{\rho}_B=\frac{1}{4}[\mathbb{I}^A\otimes\mathbb{I}^B+\sum_i a_i\sigma_i\otimes \mathbb{I}^B+\sum_i b_i\mathbb{I}^A\otimes\sigma_{j}+\sum_{i,j}a_ib_j\sigma_{i}\otimes\sigma_{j}]$. The square norm distance between $\rho^\mathrm{B}$ and $\tilde{\rho}_A\otimes\tilde{\rho}_B$ is
\begin{eqnarray}\label{trace distance}
   F&=&\mathrm{Tr}\left[(\rho^\mathrm{B}-\tilde{\rho}_A\otimes\tilde{\rho}_B)^2\right] \nonumber \\ &=&\frac{1}{4}\left[|a|^2+|b|^2+|a|^2|b|^2+
   |c|^2-2\sum_{i} c_ia_ib_i\right],
\end{eqnarray}
where $|a|^2=\sum_i a_i^2, |b|^2=\sum_i b_i^2$ and $|c|^2=\sum_i c_i^2$. Our aim is to show that $F$ admits an absolute minimum in correspondence of the product state given by the product of the marginals of $\rho^\mathrm{B}$, $\pi_{\rho^\mathrm{B}}=\rho_A\otimes\rho_B$. Setting equal to zero the derivatives of $F$ with respect to $a_i$ and $b_i$, we obtain the system ($i,j=1,2,3$)
\begin{eqnarray}\label{derivative system1}
  a_i = c_i b_i/(1+|b|^2),  \quad
  b_i = c_i a_i/(1+|a|^2).
\end{eqnarray}
From Eqs.~(\ref{derivative system1}), multiplying the first one for $a_i$, the second one for $b_i$ and summing both on the index $i$, one immediately finds $|a|=|b|$ which in turn leads to the equation for $a_i$ (and analogous for $b_i$)
\begin{eqnarray}\label{aiequation}
a_i(|a|^4+2|a|^2+1-c_i^2)=0.
\end{eqnarray}
Being $0\leq|c_i|\leq 1$ and $|a|\geq 0$, Eq.~(\ref{aiequation}) admits the only solution $a_i=0$ for each $i$ (and thus also $b_i=0$). It is now straightforward to see that the Hessian of $F$ in this critic point has nonnegative eigenvalues, so that  $F$ has a unique local minimum in $a_i=0$, $b_i=0$ which is equal to $F_\mathrm{min}=|c|^2/4$. In order to show that this is an absolute minimum, the behavior of $F$ at the border $|a|=|b|=1$ must be studied. It is easy to show that the function $F$ on the border assumes values always larger than $F_\mathrm{min}$. This means that the product state closest to a general Bell-diagonal state $\rho^\mathrm{B}$, minimizing the distance $F$, is that obtained by choosing $a_i=b_i=0$ in $\tilde{\rho}_A$ and $\tilde{\rho}_B$. This just gives the product of the marginals of $\rho^\mathrm{B}$, $\pi_{\rho^\mathrm{B}}$, being $\rho_A=\mathbb{I}^A/2=\mathrm{Tr}_B \rho^\mathrm{B}$ and $\rho_B=\mathbb{I}^B/2=\mathrm{Tr}_A \rho^\mathrm{B}$.

\end{document}